\documentclass{mn2e}

\usepackage{amsfonts}
\usepackage{amsmath}
\usepackage{graphicx}

\title[The environmental dependence of galaxy clustering]
{The environmental dependence of galaxy clustering 
       in the Sloan Digital Sky Survey}

\author[U. Abbas and R. K. Sheth]
{Ummi Abbas$^{1}$\thanks{E-mail:  ummi@phyast.pitt.edu (UA); 
                                  shethrk@physics.upenn.edu (RKS)}
 and Ravi K. Sheth$^{2}$\footnotemark[1]\\
$^{1}$Department of Physics \& Astronomy, University of Pittsburgh, 
      Pittsburgh, PA 15260, USA\\
$^{2}$Department of Physics \& Astronomy, University of Pennsylvania, 
      Philadelphia, PA 19104, USA}

\begin{document}
\pagerange{\pageref{firstpage}--\pageref{lastpage}}

\maketitle

\label{firstpage}

\begin{abstract}
A generic prediction of hierarchical clustering models is that the 
mass function of dark haloes in dense regions in the Universe should 
be top-heavy.  We provide a novel test of this prediction using a 
sample of galaxies drawn from the Sloan Digital Sky Survey.  
To perform the test, we compare measurements of galaxy clustering in 
dense and underdense regions.  We find that galaxies in dense regions 
cluster significantly more strongly than those in less dense regions.  
This is true over the entire 0.1--30~Mpc pair separation range for 
which we can make accurate measurements.  
We make similar measurements in realistic mock catalogs in which 
the only environmental effects are those which arise from the 
predicted correlation between halo mass and environment.  
We also provide an analytic halo-model based calculation of the 
effect.   Both the mock catalogs and the analytic calculation 
provide rather good descriptions of the SDSS measurements.  
Thus, our results provide strong support for hierarchical models.  
They suggest that, unless care is taken to study galaxies at fixed 
mass, correlations between galaxy properties and the surrounding 
environment are almost entirely due to more fundamental correlations 
between galaxy properties and host halo mass, and between halo mass 
and environment.  
\end{abstract}


\section{\protect\bigskip Introduction}
The correlation between galaxy properties (morphology, star 
formation rates, luminosity, color etc.) and the surrounding 
environment has been the subject of extensive studies in the last 
few decades:  dense environments are preferentially occupied by 
elliptical, red, luminous galaxies, whereas star formation rates 
are higher in less dense regions
(Dressler 1980; Butcher \& Oemler 1984; Norberg et al. 2001, 2002; 
Balogh et al. 2002; Gomez et al. 2003; Hogg et al. 2004; 
Kauffmann et al. 2004; Berlind et al. 2005; Croton et al. 2005).  
In hierarchical models, this behaviour is expected to be a 
consequence of the fact that galaxies are surrounded by dark matter 
halos, and the properties of halos (mass, formation time, 
concentration, internal angular momentum, etc.) are correlated 
with their environments (Mo \& White 1996; Sheth \& Tormen 1999, 
2002, 2004; Lemson \& Kauffmann 1999; Gottloeber et al. 2001;
Avila-Reese et al. 2005; Gao et al. 2005; 
Harker et al. 2006; Wechsler et al. 2006).  

Recently, we described how the clustering of galaxies can be 
used to test the assumption that the correlations between galaxy 
properties and their environments are {\em entirely} a consequence 
of the correlations between haloes and their environments 
(Abbas \& Sheth 2005).  
This is a strong assumption which significantly simplifies 
interpretation of the observed luminosity dependence of galaxy 
clustering (e.g. Zehavi et al. 2005).  It is also a standard 
assumption in current halo-model descriptions of galaxy 
clustering (see Cooray \& Sheth 2002 for a review).  
The main goal of this paper is to perform this test.  

This paper is arranged as follows: 
in Section 2 we show how galaxy clustering depends on environment
in the SDSS (Adelman-McCarthy et al. 2006).  In particular, we 
measure the pair correlation function in redshift space, 
$\xi(s|\delta_s)$, for a range of environments $\delta_s$, 
as well as the projected quantity, $w_p(r_p|\delta_s)$; 
the latter is free of redshift-space distortions.  
These measurements are compared with similar measurements 
in carefully constructed mock catalogs, and from an analytic 
calculation based on the halo-model.  In both the mocks and the 
analytic calculation, correlations between galaxy properties and 
environment are entirely a consequence of the correlation between 
galaxy properties and halo masses, and between halo mass and 
environment.  We summarize our results in Section 3, where we 
also discuss some implications.
An Appendix provides details of the analytical model, which 
generalizes our earlier (real-space) work so that it can be used 
to model redshift space measurements as well.  

\section{The environmental dependence of clustering}\label{measure}
To measure the environmental dependence of clustering, we must 
decide on a measure of the environment.  Abbas \& Sheth (2005) 
showed that the precise choice of environment is not particularly 
important, in the sense that different choices lead to quantitative 
but not qualitative differences.  They used $N_R$, the number of 
galaxies in a sphere of radius $R$ centred on a galaxy, as a measure 
of that galaxy's environment, and presented results for $R=5$ and 
$8h^{-1}$Mpc.  Their analysis was performed in real-space.  
Performing a similar analysis in redshift-space is complicated 
because the environmental effect we would like to test is due to 
correlations between halo masses with the real-space density.  
Therefore, we must find a definition of density in redshift space 
which is as faithful as possible to that in real-space.  

Line-of-sight redshift-space distortions can make a sphere in 
real-space appear very different in redshift space.  For instance, 
around a spherically symmetric cluster there are two main effects:  
one is due to coherent infall around the center of the cluster, which 
appears as a squashing effect along the line of sight in redshift space 
(Kaiser 1987). The second is the ``finger of God'' effect which is due 
to the virial motions of galaxies within the cluster 
(de Lapparent et al. 1986).  This shows up as an elongation of the 
cluster along the line of sight.  The squashing effect is relatively 
small, producing effects of order unity or less, whereas the 
finger-of-god distortions are more dramatic---elongations along the 
line of sight are typically about a factor of ten.  Since clusters 
have radii of a Mpc or so, fingers of god can extend up to about 
10~Mpc.  Therefore, while counts in redshift space spheres of radii 
$5h^{-1}$Mpc are not expected to faithfully trace the counts in the 
corresponding real-space spheres, counts in spheres of radii 
$8h^{-1}$Mpc, $N_8$ should be more similar.  For this reason, in what 
follows we use $N_8$ as a measure of the environment of each galaxy.  
(If we wished to push to smaller scales, we could identify all the 
fingers of god, and then ``decompress'' them, by rescaling the distances 
along the line-of-sight so that they have the same extent as across 
the line-of-sight e.g. Tegmark et al. 2004.  But performing such a 
``manicure'' is beyond the scope of the present work.)  

We use $N_8$ to divide the galaxy population into three equal-sized 
subsamples:  the third with the largest values of $N_8$ are defined 
as being the dense subsample, and the third with the smallest values 
of $N_8$ are the underdense subsample.  We then measure the correlation 
functions in these two subsamples.  

Our strategy is to make such measurements in a volume limited galaxy 
catalog, so that selection effects are minimized.  We then compare 
with similar measurements in realistic mock catalogs.  
Throughout, we show results for a flat $\Lambda$CDM model for which 
$(\Omega_{0},h,\sigma_{8}) = (0.3,0.7,0.9)$ at $z=0$.  
Here $\Omega _{0}$ is the density in units of critical density today, 
the Hubble constant at the present time is $H_{0}= 
100 h $ km s$^{-1}$ Mpc$^{-1}$, 
and $\sigma_{8}$ describes the rms fluctuations of the initial 
field, evolved to the present time using linear theory, 
when smoothed with a tophat filter of radius $8h^{-1}$~Mpc.

\subsection{The SDSS galaxy sample}\label{sdss}
We perform our analysis on a volume limited catalog extracted from 
the SDSS DR4 database (Adelman-McCarthy et al. 2006).  We chose galaxies 
brighter than  $M_r < -21$, to match the analysis of 
Zehavi et al. (2005), whose results we use below.  
The resulting catalog contains about 75000 objects with accurate 
angular positions and redshifts, where the number density is 
0.00117 ($h^{-1}$Mpc)$^{-3}$.  


\begin{figure}
  \centering
   \includegraphics[width=1.2\hsize]{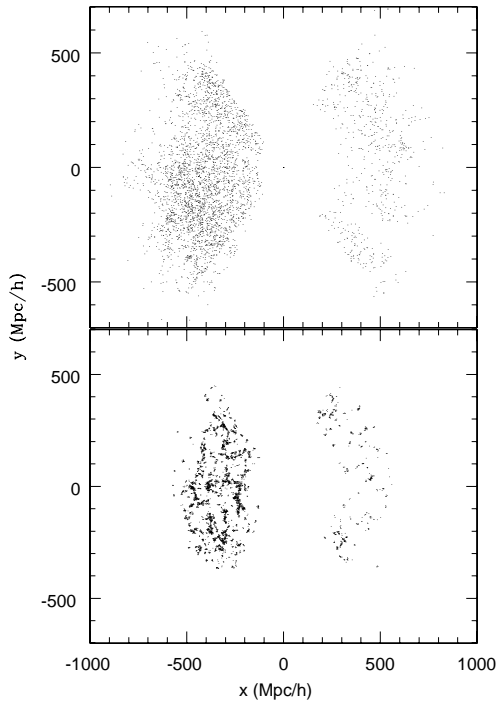}
   \caption{Pie-diagrams of the SDSS subsamples:  
             top and bottom panels show the distributions of 
             objects classified as being in the least dense 
             and the densest regions, in a slice of thickness
             100~$h^{-1}$Mpc through the survey volume.  }
   \label{sdss_dotplot}
\end{figure}

As discussed above, we define the environment of each object using 
the redshift-space information.  
Figure~\ref{sdss_dotplot} shows the spatial distribution of galaxies 
in a thin slice through the SDSS volume limited catalog.  Top panel 
shows the objects classified as being in the least dense regions, 
and bottom panel are for the objects in the densest regions.  
The galaxies in dense regions are clearly strongly clustered on 
small scales, whereas those in the underdense regions populate the 
holes defined by the spaces between the clusters that one sees in 
the dense sample.  

The following sections quantify these differences by measuring 
the correlation functions in these subsamples.  
Uncertainties on our measurements were estimated by jack-knife
resampling, in which the statistics were remeasured after omitting 
a random region, and repeated thirty times (approximately 1.5 times 
the total number of bins in separation for the results presented). 

\subsection{Mock galaxy samples}{\label{mocks}}
We have generated realistic mock galaxy samples as follows.  We start 
with the Very Large Simulation (VLS; Yoshida, Sheth \& Diaferio 2001), 
kindly made available to the public by the Virgo consortium.   
It has $512^3$ particles in a cubic box with sides $L = 479h^{-1}$Mpc.  
About 800,000 dark matter halos, each containing at least 10 particles, 
were identified in this particle distribution using the Friends-of-Friends
method. We use the simulation output for the mass, position and velocity
of each dark matter halo.

We use the results of Zehavi et al. (2005) to motivate our choice 
for how mock galaxies should be distributed within each halo.  
Specifically, to model a volume limited galaxy catalog with objects 
more luminous than $L$, halos less massive than $m_L$ are assumed to 
contain no galaxies; $m_L$ depends on the galaxy population under 
consideration.  Galaxies more massive than $m_L$, contain one central 
galaxy, and may also contain satellite galaxies.  The number of 
satellites is drawn from a Poisson distribution with mean 
$\langle N_s|m\rangle$, where
\begin{equation}
	\langle N_s|m\rangle = \left(\frac{m}{m_1}\right)^\alpha \qquad 
                  {\rm if} \,\,m \geq m_L.
 \label{NsatM}
\end{equation}
For SDSS galaxies more luminous than $M_r < -21$, 
$m_L = 10^{12.72} h^{-1}M_{\odot}$, $m_1 = 23 m_L$, and, 
$\alpha = 1.39$ (Zehavi et al. 2005).  
(A Poisson distribution for the number of satellites is motivated by 
the work of Kravtsov et al. 2004).  
We then assume that the satellites in a halo are distributed 
around the halo center similarly to the dark matter 
(e.g. Navarro et al. 1997).  

To model redshift space effects, we must model the velocity vector 
of each mock galaxy.  We do so by assuming that 
 $v_{\rm gal} = v_{\rm halo} + v_{\rm vir}$, 
where $v_{\rm halo}$ is the halo motion provided by the simulation, 
and $v_{\rm vir}$ is obtained as follows.  The central galaxy in a 
halo is assumed to be at rest with respect to the halo, so $v_{vir}=0$.   
The virial motions of satellite galaxies are modelled by assuming 
that haloes are isotropic, virialized, and isothermal with Maxwellian 
velocities around the halo center.  The one-dimensional velocity 
dispersion is 
$1000\,(r_{200}h/{\rm Mpc})/\sqrt{2}$, where $r_{200}$ is 
the scale on which the enclosed mass is 200 times the critical density:  
 $m = 200\bar\rho_{\rm crit}\,(4\pi r_{200}^3/3)$.  
Following results in Sheth \& Diaferio (2001), we assume that this 
virial term is independent of local environment.  

In the distant observer approximation, the position in redshift space 
is 
	$s = x + v_x/H_0$,
where $x$ is the real-space coordinate in the $x$-direction, 
$v_x$ is the $x$-component of the peculiar velocity.  
and $s$ is the redshift-space distance in the $x$-direction.  
The $y$- and $z$- components of the position are unchanged.  
The isothermal Maxwellian assumption means that the virial motions 
add Gaussian noise to the line-of-sight position of each satellite 
galaxy.  

\begin{figure}
 \centering
 \includegraphics[width=\hsize]{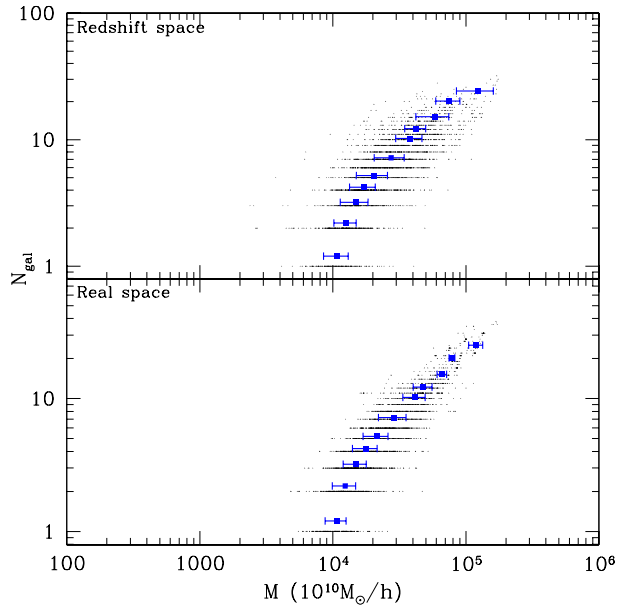}
 \caption{Comparison of local density estimates within real (bottom) 
          and redshift (top) space spheres of radius $8h^{-1}$Mpc.
          The median (given by the squares) and quartile range of 
          halo mass corresponding to certain number of galaxies is 
          shown (for clarity these points have been shifted upwards).}
 \label{N8scatter}
\end{figure}

We then measure $N_8$ for each galaxy by counting the total 
number of galaxies within $8h^{-1}$Mpc.  For the mock catalog, 
we can do this in both real- and redshift-space.  
Figure~\ref{N8scatter} compares these two estimates of the 
local density.  They are not widely different, suggesting 
that the analysis in the Appendix will be useful.  

\begin{figure}
 \centering
 \includegraphics[width=\hsize]{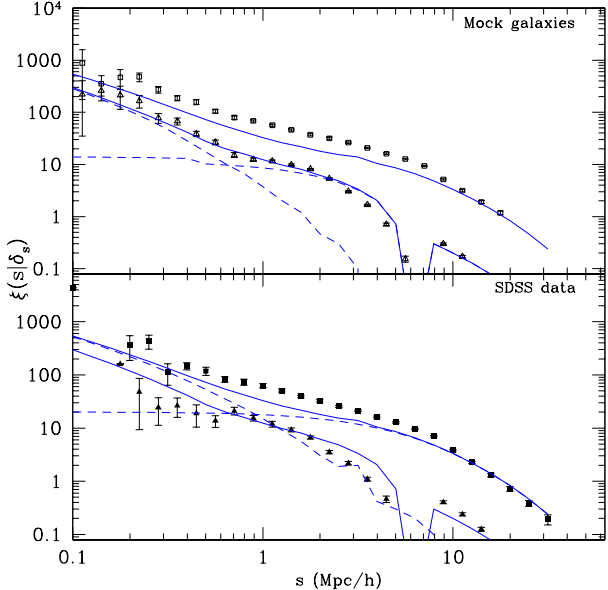}
 \caption{Environmental dependence of the galaxy correlation function 
   in redshift space. Upper panel shows measurements in the mock catalog, 
   and lower panel shows measurements in the SDSS.  
   In both cases, the galaxy catalog is volume limited to $M_r<-21$,  
   and the environment of a galaxy was defined by counting the 
   number of galaxies within a redshift-space sphere of radius 
   $8h^{-1}$Mpc centred on it. 
   The squares and triangles in each panel show $\xi(s|\delta_s)$ 
   for the galaxies in the densest 1/3 and least dense 1/3 of the 
   sample.  
   Solid curves show the analytic model for $\xi(s|\delta_s)$ 
   that is developed in Appendix~\ref{model}.
   Dashed curves in the upper panel show the 1-halo (dominates on 
   small scales) and the sum of the 2h-1p and 2h-2p contributions 
   (dominate on intermediate and large scales, respectively) to 
   $\xi(s|\delta_s)$ of the less-dense sample.  
   In the lower panel, the dashed curves show these 
   contributions for the dense sample.  }
 \label{zvls}
\end{figure}

\subsection{Results}

Figure~\ref{zvls} quantifies the spatial differences seen in 
Figure~\ref{sdss_dotplot}; it shows the redshift 
space correlation functions in dense and underdense regions measured 
in the mock catalogs (upper panel) and in the SDSS volume limited 
catalog (lower panel). 
In both panels, $\xi(s|\delta_s)$ for the dense sample is 
significantly larger than it is in the underdense sample.  
On large scales, this is because dense regions host the most massive 
haloes which in turn contain many galaxies; on smaller scales, the 
fact that the halo density profiles depend on halo mass also matters 
(Abbas \& Sheth 2005). 
The inflection or break at the scale on which we define the
environment (8$h^{-1}$ Mpc), which is seen in the clustering signal
for underdense regions, arises because this scale is significantly
larger than the virial radius of a typical halo.
Let $R$ denote the scale on which the environment is defined.
Then, pairs which come from different halos are of two types:
those separated by scales smaller than $R$ are said to be in the
same patch, whereas more widely separated pairs are in different
patches.  Abbas \& Sheth (2005) called these the 2$h-1p$ and $2h-2p$ 
contributions to the statistic.
Now, by definition, there are no $2h-2p$ pairs with separations
smaller than 8$h^{-1}$ Mpc, so $\xi_{2h-2p} = -1$ on smaller scales.
In addition, underdense regions are those with small $N_R$, so they
have few pairs in the $2h-1p$ term by definition.
In the limit in which there is only one halo in each underdense
patch (i.e., the one surrounding the galaxy around which the patch 
was centered), there will be no $2h-1p$ pairs.  In this limit, the 
correlation function is the sum of the 1h term, which falls rapidly 
on scales larger than the virial radius (a few Mpc) and the $2h-2p$ 
term (which is only significant on scales larger than the patch 
radius).  Therefore, in this limit, if $R$ is significantly larger 
than the virial radius of a typical halo, there will be a dramatic 
feature in $\xi$ at scale $R$.  As the number of $2h-1p$ pairs 
increases, this feature becomes less obvious.  Indeed, in dense 
regions---those which have larger $N_R$ and so have more $2h-1p$ 
pairs, there is little evidence of this feature.

\begin{figure}
 \centering
 \includegraphics[width=\hsize]{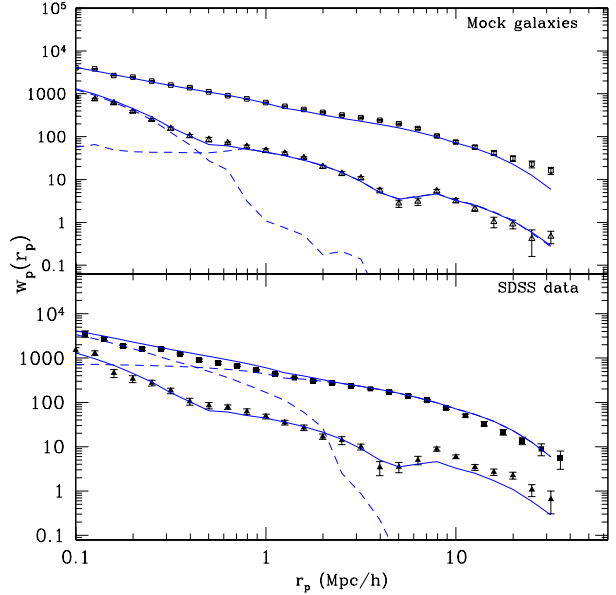}
 \caption{Similar to Figure~\ref{zvls}, but now for the projected galaxy  
 correlation function. The upper panel shows measurements in the mock 
 catalog, and the lower panel is for the SDSS.  
 Symbols and line-styles are the same as for Figure~\ref{zvls}.}
 \label{wprp}
\end{figure}

The solid curves in the two panels show the analytic calculation 
outlined in the Appendix.  They provide a reasonable description 
of the measurements in both panels.  However, while the agreement 
is good on large scales, the curves underestimate the small-scale 
signal in dense regions.  Since these smaller scales are the ones
most affected by finger-of-god distortions, it may be that the 
discrepancy is due to inadequacies in the analytic treatment of 
redshift-space effects (see Scoccimarro 2004 for a discussion of 
the sorts of effects our analysis ignores).  

To eliminate this source of uncertainty, we have also studied 
the projected quantity 
\begin{equation}
	w_p(r_p|\delta_s) = 2 \int_{0}^{\infty} d\pi\,\xi(r_p,\pi|\delta_s);
\end{equation}
where $r=\sqrt{r_p^2 + \pi^2}$. We integrate up to $\pi$ = 35 $h^{-1}$Mpc, 
which is large enough to include most correlated pairs. 
Figure~\ref{wprp} shows the results, both in the mock catalog (top 
panel) and in the SDSS (bottom panel).  Now, the agreement with the 
analytical model is very good, suggesting that our analytic treatment 
of redshift-space distortions is inadequate. Once again, the inflection
at the scale of the patch size for the underdense sample
is caused due to the transition from one type of 2-halo term to the other.

Both for $\xi(s|\delta_s)$ and $w_p(r_p|\delta_s)$ the differences 
between the two environments are dramatic---they are measured with 
high statistical significance.  
Nevertheless, the analytic model, which only incorporates those 
correlations with environment which arise from the correlation 
between halo mass and environment, provides an excellent description 
of the measurements.  This leaves little room for other environmental 
effects.

\section{Discussion and Conclusions}
One of the luxuries of the latest generation of large-scale sky 
surveys is that they contain sufficiently many objects that one 
can study subsamples of galaxies divided up in various ways.  
Here, we have focused on the clustering of galaxies in a volume 
limited sample drawn from the SDSS, and studied how the 
clustering of these galaxies depends on environment.  
We find that galaxies in dense regions are considerably more strongly 
clustered than those in less dense regions (Figures~\ref{zvls} 
and~\ref{wprp}).  

This is perhaps not so surprising---after all, a dense region is 
one in which many galaxies are crowded together.  
What is more surprising is that this dependence on environment is 
very well reproduced by numerical (Section~\ref{mocks}) and analytic 
(Appendix~A) models in which the entire effect is due to the fact 
that galaxy properties correlate with the masses of their parent 
halos, and massive halos preferentially populate dense regions.  
Hierarchical models make quantitative predictions for this correlation 
between halo mass and environment, and so the agreement between our 
models and the measurements provides strong support for such models.  
In this respect, our results are consistent with those of 
Mo et al. (2004), Kauffmann et al. (2004), 
Berlind et al. (2005), Blanton et al. (2006)
and Skibba et al. (2006); this is reassuring, since our methods 
are very different.  

Our test of environmental effects is particularly interesting in 
view of recent work showing that, at fixed mass, haloes in dense 
regions form earlier (Sheth \& Tormen 2004), and that this effect 
is stronger for low mass haloes (Gao et al. 2005; Harker et al. 2006; 
Wechsler et al. 2006).  Such a correlation is not part of our 
analytic model, nor is it included in our mock catalogs.  Presumably, 
the good agreement with the SDSS measurements is due to the fact that 
we have concentrated on luminous galaxies, and these populate the 
more massive haloes.  It will be interesting to see if this agreement 
persists at lower luminosities.  

The agreement between our models and the measurements has an 
important consequence:  Unless care is taken to study a population 
at fixed halo mass, our results indicate that observed correlations 
between gastrophysical effects (e.g. ram pressure stripping, 
strangulation, harrassment) and environment are dominated by the fact 
that these effects also correlate with halo mass, and halo mass 
correlates with environment.  

Larger samples will allow us to study if these trends persist to 
fainter, presumably less massive galaxies.  
And more distant samples will allow us to study if these trends 
evolve.

\section*{Acknowledgments}

We thank Ramin Skibba for many helpful discussions, 
Cameron McBride and Jeff Gardner for the NTropy code which was 
used to measure the correlation functions and projected statistics 
in the simulations and data, 
Andrew Connolly and Ryan Scranton for providing the SDSS data 
samples used in this paper, and the Virgo consortium for making 
their simulations available to the public.
We also thank the referee for suggesting changes that helped to 
improve the paper.

Funding for the Sloan Digital Sky Survey (SDSS) has been provided by the
Alfred P. Sloan Foundation, the Participating Institutions, the National
Aeronautics and Space Administration, the National Science Foundation, the U.S.
Department of Energy, the Japanese Monbukagakusho, and the Max Planck Society.
The SDSS Web site is http://www.sdss.org/.

The SDSS is managed by the Astrophysical Research Consortium (ARC) for the
Participating Institutions. The Participating Institutions are The University
of Chicago, Fermilab, the Institute for Advanced Study, the Japan Participation
Group, The Johns Hopkins University, the Korean Scientist Group, Los Alamos
National Laboratory, the Max-Planck-Institute for Astronomy (MPIA), the
Max-Planck-Institute for Astrophysics (MPA), New Mexico State University,
University of Pittsburgh, University of Portsmouth, Princeton University, the
United States Naval Observatory, and the University of Washington.

\appendix

\section{The Analytical model}{\label{model}}


This Appendix discusses how the halo model calculation of environmental 
effects on clustering can be extended to include redshift-space effects.  
Our strategy is to combine the halo-model description of redshift space 
effects (White 2001; Seljak 2001) with the halo model description of 
environmental effects provided by Abbas \& Sheth (2005).  

In redshift space, two effects modify the real space expressions 
derived by Abbas \& Sheth (2005). One of these is a boost of power on 
large scales due to the instreaming of matter into overdense regions 
(Kaiser 1987); this affects the 2-halo terms.  
Using density conservation to linear order and making the distant 
observer approximation, the redshift-space galaxy density perturbation 
can be written as 
\begin{equation}
  \delta_{g}^{rs} = \delta_{g} + \delta_v\mu^2 
  \label{dskaiser}
\end{equation}
where $\mu = \hat{r}\cdot \hat{k}$, $\delta_g$ is the real space 
galaxy density perturbation and $\delta_v$ is the velocity divergence. 
This is related to the density perturbation $\delta_{dm}$ via 
$\delta_v = f \delta_{dm}$, where 
 $f(\Omega) \equiv d \log \delta/d \log a \simeq \Omega^{0.6}$, 
and $a$ is the scale factor. 

The other effect is the suppression of power due to the virial motions 
within haloes; this affects the 1-halo term (Sheth 1996). 
The assumption of isotropic, isothermal, Maxwellian motions within 
halos means that the effect can be modeled as a convolution with a 
Gaussian.  In particular, the density contrast in redshift space is
\begin{equation}
 \delta_g^{rs} = \delta_{g} e^{-(k \sigma \mu)^2/2}.
 \label{dsvir}
\end{equation}
Scoccimarro (2004) discusses why these descriptions 
(equations~\ref{dskaiser} and~\ref{dsvir}) of redshift-space 
distortions are rather crude.  For our purposes, they represent 
reasonable first approximations to a more sophisticated model.  

Let $n(M,V)$ denote the number density of patches of mass $M$ and volume
$V$, and let $N(m|M,V)$ be the average number of $m$ haloes in regions 
of volume $V$ which contain mass $M$.
The isotropized redshift space power spectrum is obtained by averaging 
$(\delta_g^{rs})^2$ over $\mu$, $m$ and $M$.  
In particular, the 1-halo term can be written as,
\begin{equation}
 \begin{split}
   P&_{1h}^{gal}(k|\delta) =  \int_{M_{min}}^{M_{max}} dM\, n(M,V) 
        \int_0^M dm \,N(m|M,V)\,\\ 
        & \times\ 
          {[\langle 2 N_s|m\rangle\,u(k|m)\Re_1(k\sigma) 
	      +\langle N_s|m\rangle^2\,|u(k|m)|^2\Re_2(k\sigma)]
            \over\bar n_{\delta-gal}^2},
 \end{split}
\end{equation}
where
\begin{equation}
  \Re_p(\alpha = k \sigma [p/2]^{1/2}) = \frac{\sqrt{\pi}}{2}
  \frac{{\rm erf}(\alpha)}{\alpha}
\end{equation}
for $p = 1,2$, 
and $\bar n_{\delta-gal}$ is the number density of galaxies surrounded 
by regions containing at least $N_{min}$ other galaxies:  
\begin{equation}
 \begin{split}
 \bar n_{\delta-gal} &= \int_{M_{min}(N_{min})}^{M_{max}(N_{max})}
      \!\!\!\! dM\,n(M,V)\\
      \ & \qquad\times\quad  \int_0^M dm\,N(m|M,V)\,\langle N_{gal}|m\rangle.
 \end{split}
\end{equation}
Here, $\langle N_{gal}|m\rangle = 1 + \langle N_s|m\rangle$ is the 
average number of galaxies occupying a halo of mass $m$ (in our 
model, it is zero below some $m_L$; c.f. equation~\ref{NsatM}).   
In practice, $M_{min}(N_{min})$ is obtained by varying $N_{min}$ 
until the value of this expression matches the observed number 
density.  

The two-halo term is more complex as it now has two types of 
contributions: pairs which are in the same patch (2h-1p),
and pairs in different patches (2h-2p).  
The 2h-1p term can only be important on intermediate scales 
(i.e., those which are larger than the diameter of a typical halo but 
smaller than the diameter of a patch).  
The 2h-2p term is 
\begin{equation}
  P_{2h-2p}^{gal} = (F_g^2 + \frac{2}{3}F_v F_g + \frac{1}{5}F_v^2) 
  P_{\rm Lin}(k|R_p),
\end{equation}
where
\begin{equation}
  \begin{split}
    F_v &= f  \int_{M_{min}}^{M_{max}} dM\,n(M,V)\,B(M,V) \\ 
  \ & \,\,\, \times \int_0^M dm\,N(m|M,V)\,\frac{m}{\rho_\delta} 
   \Re_1(k \sigma) u(k|m) \\
    F_g &= \int_{M_{min}}^{M_{max}} dM\,n(M,V)\,B(M,V) \\ 
  \ & \,\,\, \times\ \int_0^M dm\,N(m|M,V)\,\frac{\langle N_{gal}|m\rangle}
    {\bar n_{\delta-gal}}\Re_1(k \sigma) u(k|m);
  \end{split}
\end{equation}
$P_{\rm Lin}(k|R_p)$ denotes the power spectrum associated 
with setting the linear theory correlation function to $-1$ on 
scales smaller than the diameter of a patch $2R_p$.  This 
truncation has little effect on small $kR_p\ll 1$, where 
$P_{\rm Lin}(k|R_p)\approx P_{\rm Lin}(k)$.  
And the factor $B(M,V)$ describes the bias associated with the 
clustering of the patches; it depends on the abundance of such 
patches (see Abbas \& Sheth 2005 for details).

Similarly, the 2h-1p term can be written as
\begin{equation}
  \begin{split}
  P_{2h-1p}^{gal}(k|\delta) &= 
  \int_{M_{min}}^{M_{max}} dM \, n(M,V)(F_g^{'2} + \frac{2}{3}F'_v F'_g + 
  \frac{1}{5}F_v^{'2}) \\
  \ & \qquad\qquad \times \Bigl[P_{Lin}(k)- P_{Lin}(k|R_p)\Bigr],
  \end{split}
\end{equation}
where $P_{\rm Lin}(k)-P_{Lin}(k|R_p)$ denotes the power spectrum 
associated with setting the linear theory correlation function to 
zero on scales larger than the diameter of a patch $2R_p$, and 
\begin{equation}
  \begin{split}
    F'_v &= f \int_0^M dm\,N(m|M,V)\,\frac{m}{\rho_\delta} 
   \Re_1(k \sigma) b(m)\, u(k|m) \\
    F'_g &= \int_0^M dm\,N(m|M,V)\,\frac{\langle N_{gal}|m\rangle}
    {\bar n_{\delta-gal}}\Re_1(k \sigma) b(m) \, u(k|m).
  \end{split}
\end{equation}
Here $b(m)$ is the bias factor of haloes of mass $m$ (from 
Sheth \& Tormen 1999). 
The correlation function, $\xi(s)$, is obtained by taking the 
Fourier transform of the power spectrum $P(k)$.

\label{lastpage}

\end{document}